\documentstyle[12pt]{article}

\begin{document}

\begin{titlepage}
\setcounter{page}{1}

\author{Waldemar Puszkarz\thanks{
Electronic address: puszkarz@cosm.sc.edu} \\
{\small {\it Department of Physics and Astronomy,} }\\
{\small {\it University of South Carolina,} }\\
{\small {\it Columbia, SC 29208}}}
\title{{\bf Is an Electromagnetic Extension of the Schr\"{o}dinger Equation
Possible?}}
\date{{\small (October 8, 1997)}}
\maketitle

\begin{abstract}
The idea of equivalence of the free electromagnetic phase and 
quantum-mechanical one is investigated in an attempt to seek modifications
of Schr\"{o}dinger's equation that could realize it. It is assumed that
physically valid realizations are compatibile with the $U(1)$-gauge and
Galilean invariance. It is shown that such extensions of the Schr\"{o}dinger
equation do not exist, which also means that despite their apparent
similarity the quantum-mechanical phase is essentially different from the
electromagnetic one.

\vskip 0.5cm \noindent 
\end{abstract}

\end{titlepage}


The fundamental equation of wave mechanics was put forward by Erwin
Schr\"{o}dinger 70 years ago. Ever since it has been subjected to a thorough
theoretical and experimental examination. As it was soon demonstrated 
by Madelung, this decent linear equation,
\begin{equation}
i\frac{\partial \Psi }{\partial t}=-\frac{1}{2m}\Delta \Psi +V\Psi  \label{1}
\end{equation}
can also be formulated as a system of two, highly nonlinear equations for
the phase $S$ and the amplitude $R$ of the wave function $\Psi =R\exp(iS)$,
\begin{equation}
\frac{\partial R^{2}}{\partial t}+\frac{1}{m}\vec{\nabla}\cdot \left( R^{2}%
\vec{\nabla}S\right) =0,  \label{2}
\end{equation}
\begin{equation}
\frac{1}{m}\Delta R-2R\frac{\partial S}{\partial t}-2RV-\frac{1}{m}R\left( 
\vec{\nabla}S\right) ^{2}=0.  \label{3}
\end{equation}
One obtains this system by taking the imaginary and the real part of (1) or by
deriving it from the Lagrangian for the Schr\"{o}dinger equation 
\begin{equation}
L^{SE}({\vec{r}},t)=\frac{i}{2}\left( \Psi ^{*}\frac{\partial \Psi }{%
\partial t}-\frac{\partial \Psi ^{*}}{\partial t}\Psi \right) -\frac{1}{2m}%
\vec{\nabla}\Psi ^{*}\vec{\nabla}\Psi +\Psi ^{*}V\Psi .  \label{4}
\end{equation}
which can also be cast in terms of $R$ and $S$ as
\begin{equation}
L^{SE}(R,S)=R^{2}\frac{\partial S}{\partial t}+\frac{1}{2m}\left[ \left( 
\vec{\nabla}R\right) ^{2}+R^{2}\left( \vec{\nabla}S\right) ^{2}\right]
+R^{2}V.  \label{5}
\end{equation}
Above and throughout the rest of this paper we use the system of natural 
units $\hbar=c=1$.

The Madelung formulation (2-3) of the Schr\"{o}dinger equation is
perhaps the best demonstration of this equation hidden nonlinear features.
Moreover, the fact that it employes the most natural variables the standard
interpretation of quantum mechanics offers seems to be adding extra strength
to it. According to the probabilistic interpretation of quantum mechanics, 
$\rho =R^{2}$ can be thought of as the density probability of a quantum
system. The equation (2) is then interpreted as the continuity equation with
the probability current revealed to be $j=\frac{1}{m}R^{2}\vec{\nabla}S$.
Considering that the Madelung formulation was discovered in the same year as
the Schr\"{o}dinger equation, it may seem surprising that attempts at
nonlinear modification of this equation came much later. The best known of
them, the cubic Schr\"{o}dinger equation, modifies the original
Schr\"{o}dinger equation by supplementing the RHS of (1) with the term $%
|\Psi|^2\Psi$. It turns out that this modification fails to preserve the
separability of noninteracting systems \cite{Bial}. This assumption, a
rather natural feature of Schr\"{o}dinger's equation is perhaps the only
physical criterion put forward so far that a nonlinear modification of this
equation should satisfy. In fact, one can divide such modifications into
those that meet it and the remaining ones. Since it is rather nontrivial to
find a modification that satisfies this assumption, it appears that there is
only a limited class of modifications that fall into the first category [1,
2, 3], although a general modification scheme of Weinberg \cite{Wein} does 
provide a class of specific modifications of its own. To justify the 
modifications from the other category one needs to provide arguments of a 
different kind, be it of simplicity or integrability as is the case for the 
cubic Schr\"{o}dinger equation or of some particular or desirable properties. 
One of the most notable proposals in this class seems to be the modification
suggested by Staruszkiewicz \cite{Sta1}, recently extended by this author 
\cite{Pusz}.

The Staruszkiewicz modification was originally meant as an electromagnetic
generalization of the Schr\"{o}dinger equation based on his theory of the
free electromagnetic phase [6,7]. Coming from this theory is the term that
modifies the Schr\"{o}dinger equation rendering its Lagrangian to be 
\begin{equation}
L^{SE}_{mod}=R^2 \frac{\partial S}{\partial t} + \frac{1}{2m}%
\left[\left(\vec\nabla R\right)^2 +R^2 \left(\vec\nabla S\right)^2\right]
+R^2V+ 2\gamma\left(\Delta S\right)^2 .  \label{6}
\end{equation}
The main and unique feature of Staruszkiewicz's modification is the
dimensionless character of the constant $\gamma$ in the system of natural
units, $\hbar=c=1$, in three dimensions. All other modifications introduce
some dimensional constants. It was originally hoped that the term $%
\left(\Delta S\right)^2$ was also unique for three dimensions. This is
however not the case. It is possible to extend the Staruszkiewicz
modification in a way that preserves this unique property of it, but results
in six modification terms altogether \cite{Pusz}. As a matter of fact, such
an abundance of terms appears to be a generic feature of the modifications
of this fundamental equation.

In this historical approach\footnote[1]{We call this approach ``historical'' 
after Professor A. Staruszkiewicz in a private conversation with this author. 
The reason this approach needs to be revised is the modification term that 
is not invariant under the $U(1)$-gauge transformation as opposed to the 
original Schr\"{o}dinger equation.
Since there is no neccessity to sacrifice this invariance one should abandon
the assumption (7) and treat the quantum-mechanical phase $S$ as it is
without identifying it with the electromagnetic one.} incorporated was in an
essential manner the assumption that the electromagnetic phase and the
quantum-mechanical one are proportional to each other, the proportionality
in question involving the electronic charge according to the following
simple formula 
\begin{equation}
S_{elm}=\frac{1}{e}S_{quantum}.  \label{7}
\end{equation}
We will call this last equation the assumption of the equivalence of the
electromagnetic $S_{elm}$ and quantum-mechanical $S_{quantum}$ phase. For
simplicity, we will omit the subscripts whenever it does not lead to a
confusion.

It is the main purpose of the present paper to examine whether it is
possible to construct modifications of the Schr\"{o}dinger equation that
would incorporate this assumption in a physically sensible manner, that is,
in the first place, without compromising fundamental physical principles. In
particular, this means that we will insist on preserving both the $U(1)$%
-gauge invariance of the Schr\"{o}dinger equation and its Galilean
invariance. We will refer to them as electromagnetic extensions of the
Schr\"{o}dinger equation.

The modifications we intend to construct are very peculiar in this respect
that they are based on a theory whose character is certainly less
fundamental than that of quantum theory. Let us notice that $S_{elm}$ is the 
$U(1)$ phase related to electromagnetism and having an intimate relation to
the electronic charge which is the generator of symmetries that preserve the
structure of Maxwell equations. It is therefore not trivial that in (6) the
same phase is promoted to the quantum-mechanical phase. The two phases, {\it %
apriori} completely different as representing different physical entities,
can only formally be related via (7).

The difference between these two objects becomes clear when one realizes
that $S_{quantum}$ appears only in the combination with $R$ and is subjected
to the nonlinear Schr\"{o}dinger equation (2-3), while the electromagnetic
phase is a loose cannon whose evolution is not determined by any generic
physical equation. Only in a model that breaks the gauge invariance one is
able to work out the equation of motion for it. In any model that respects
this invariance the electromagnetic phase can be freely chosen. The very
essence of the intended modifications is to postulate that this phase
represents the same physical reality as the quantum phase. It is certainly
not obvious though that these phases can be identified in the above sense
and it is particularly difficult to insist on this for systems whose nature
is by no means electromagnetic. Only if one deals with electromagnetic
systems can one venture such an assumption.

Let us assume that we deal with such systems. One is then required to modify
the Schr\"{o}dinger equation so that it takes into account the presence of
electromagnetic potentials $\vec A =\frac{1}{e}\vec \nabla S$ and $\Phi=%
\frac{1}{e} \frac{\partial S}{\partial t}$. Similarly as in the original
Staruszkiewicz modification, we choose the electromagnetic potentials to be
pure gauge to ensure a complete quantum character of our modification. This
choice guarantees that the electromagnetic fields are absent and the only
way the behavior of a quantum system is affected can originate from the
potentials themselves, as in the celebrated Aharonov-Bohm effect \cite{Aha}.
This is unlike in classical mechanics where pure-gauge potentials have no
impact on the behavior of a charged particle.

The Schr\"{o}dinger equation for a spinless particle carrying a charge $e$
in the minimal coupling scheme reads 
\begin{equation}
i\frac{\partial\Psi}{\partial t}=-\frac{1}{2m}(\vec\nabla-ie\vec A)^2\Psi +
e\Phi\Psi + V\Psi.  \label{8}
\end{equation}
By taking the real and imaginary part of the last equation one obtains 
\begin{equation}
\frac{1}{2m}\Delta R-2R\frac{\partial S}{\partial t}-R V=0  \label{9}
\end{equation}
and 
\begin{equation}
\frac{\partial}{\partial t}R^2=0.  \label{10}
\end{equation}

We have not attempted to introduce any new terms as we want to show that
even without them the assumed equivalence of the phases runs into
difficulties. It turns out that the implementation of (7) along with the
potentials of the form $A_{\mu}=\frac{1}{e}\partial_{\mu}S$ results in
equations that do not respect the Galilean invariance.

This is seen from the last equations if we recall that under the Galilean
transformation $t=t^{\prime},\,\,\, \vec x=\vec x^{\prime}+\vec v t$\, the
phase $S$ changes as 
\begin{equation}
S(t,\vec x) =S^{\prime}(t^{\prime},\vec x^{\prime}) + m\vec v\cdot \vec
x^{\prime}+ \frac{m}{2}\vec v^2 t^{\prime}  \label{11}
\end{equation}
while 
\begin{equation}
\nabla=\nabla^{\prime}, \,\,\, \frac{\partial}{\partial t}=\frac{\partial}{%
\partial t^{\prime}}- \vec v \cdot \vec \nabla^{\prime}.  \label{12}
\end{equation}
Under the same nonrelativistic transformation of coordinates the components
of the four-potential $A_{\mu}=(\Phi/c, \vec A)$ transform as 
\begin{equation}
\Phi^{\prime}= \Phi-\vec v \cdot\vec A , \,\,\, \vec A^{\prime}=\vec A.
\label{13}
\end{equation}
Now, substituting $A_{\mu}=\frac{1}{e}\partial_{\mu}S$ into the RHS
of the last equations and working them out in terms of $S^{\prime}$, using  
(11) and (12), one obtains the results that do not coincide with the LHS of 
equations (13). For example, the RHS of the first of the equations concerned 
leads to 
$\frac{1}{e}\left(\frac{\partial S^{\prime}}{\partial t^{\prime}}-
2\vec v \cdot \vec\nabla^{\prime}S^{\prime}-\frac{3}{2}mv^2 \right)$ 
that is simply not the same as 
$\frac{1}{e}\frac{\partial S^{\prime}} {\partial t^{\prime}}$ which
one obtains on the other side of this equation. Clearly, this is the reason
why the Galilean invariance fails in this particular case.

We have therefore demonstrated that the assumption of the equivalence of
phases cannot be used to construct a modification of the Schr\"{o}dinger
equation that would be consistent with both the $U(1)$ gauge and Galilean
invariance. In a broader sense, what we have shown by this is that the
quantum-mechanical phase is essentially different from the electromagnetic
one despite their formal similarity.

\section*{Acknowledgments}

I would like to thank Professor Pawe{\l} O. Mazur for introducing me to the
work of Professor Starusz-\newline kiewicz. I am also indebted to 
Professor Andrzej Staruszkiewicz for a discussion of his modification. 
This work was partially supported by the NSF grant No. 13020 F167 and 
the ONR grant R\&T No. 3124141.

\end{document}